\documentclass[aps,preprint,showpacs]{revtex4}
\usepackage{graphicx}
\usepackage{url}
\bibstyle{apsrev}

\newcommand{\ky}{k_y\rho_i}
\newcommand{\lte}{L_{T_e}}
\newcommand{\vthi}{v_{{\rm th},i}}
\newcommand{\vthe}{v_{{\rm th},e}}
\newcommand{\lcorr}{L_c}
\newcommand{\dmag}{D_m}

\begin{document}

\title{Microtearing Modes in Reversed Field Pinch Plasmas}

\author{I. Predebon, F. Sattin, M. Veranda, D. Bonfiglio, S. Cappello}

\affiliation{Consorzio RFX, Associazione EURATOM-ENEA sulla fusione, Padova, Italy}

\date{\today}

\begin{abstract}
In the reversed field pinch RFX-mod strong electron temperature gradients develop when the Single-Helical-Axis regime is achieved. Gyrokinetic calculations show that in the region of the strong temperature gradients microtearing instabilities are the dominant turbulent mechanism acting on the ion Larmor radius scale. The quasi-linear evaluation of the electron thermal conductivity is in good agreement with the experimental estimates.
\end{abstract}


\maketitle


Transport barriers in magnetically confined plasmas represent one of the topical subjects of fusion science, whose ambitious aim is the controlled release of thermonuclear energy in future power plants. Barriers~\cite{itbs}, i.e., regions where the transport of particles and energy is substantially reduced, are phenomena characterized by complexity, multi-scale interactions and self-organization processes. Understanding them is a challenge for both theory and modelling, also due to the variety of magnetic confinement devices - tokamaks, stellarators and now reversed-field pinches (RFPs) - where they are encountered. In this Letter we identify, for the first time, a key mechanism driving the physics of transport in the barriers observed in the RFP device RFX-mod~\cite{rfx}. When the self-organized Single-Helical-Axis (SHAx) regimes are achieved~\cite{shax}, the plasma approaches equilibria with conserved helical magnetic surfaces, normally endowed with internal transport barriers (ITBs) which bind the hot core~\cite{eps09}. This new experimental condition raises the question whether small-scale transport mechanisms, until now overshadowed by the large-scale MHD turbulence, may rule the fundamental properties of the RFP plasma.

The strong gradients typical of ITBs are reservoirs of large amounts of free energy available to trigger microinstabilities. Their effect is to increase radial transport and damp a further increase of the gradients. In tokamaks, this has been a main topic of research for several decades. Conversely, studies applied to the RFP configuration started only recently, mainly focussed on electrostatic Ion Temperature Gradient (ITG) modes. In this respect, various approaches agree that present-day RFPs are subcritical to such instabilities, due to the strong Landau damping of the modes~\cite{itgs,gs2itg}.

An instability that attracted the interest of the tokamak community in the '70s is the MicroTearing (MT) mode~\cite{drake77}. The MT mode is a high-wavenumber drift-tearing mode driven unstable by the electron temperature gradient, unlike its long-wavelength counterpart which is essentially current-driven. The existence of MT modes leads to chains of magnetic islands, whose overlapping brings to local stochastization of magnetic field lines. Hence, it was speculated that MT modes might provide an effective contribution to the thermal diffusivity through electron parallel motion along the wandering field lines. However, these modes are expected to become stable at relatively low collisionality, $\nu\ll\omega^*$, with $\nu$ electron collision frequency and $\omega^*$ electron diamagnetic frequency~\cite{gladd80}. These conditions are usually met in current high-temperature devices, which are effectively collisionless in the core. For this reason, in tokamaks, MT modes are nowadays studied either in connection with edge transport~\cite{edgemt}, or in the core of medium-temperature devices~\cite{sphtok}. Quite recently, however, some broader interest in core MT modes has revived in connection with strong ITBs and their possible coupling with other microinstabilities, like the Electron Temperature Gradient (ETG) modes~\cite{horton10}. The ohmically-heated RFX-mod presently features a moderate peak temperature (1 to 1.5 keV) which, coupled to the strong internal temperature gradients, should represent an optimal environment for these modes to grow. Given the long-established importance of tearing instabilities in the RFP, we found it natural to consider the connection between MT modes and ITBs in this machine.

\begin{figure}
\includegraphics{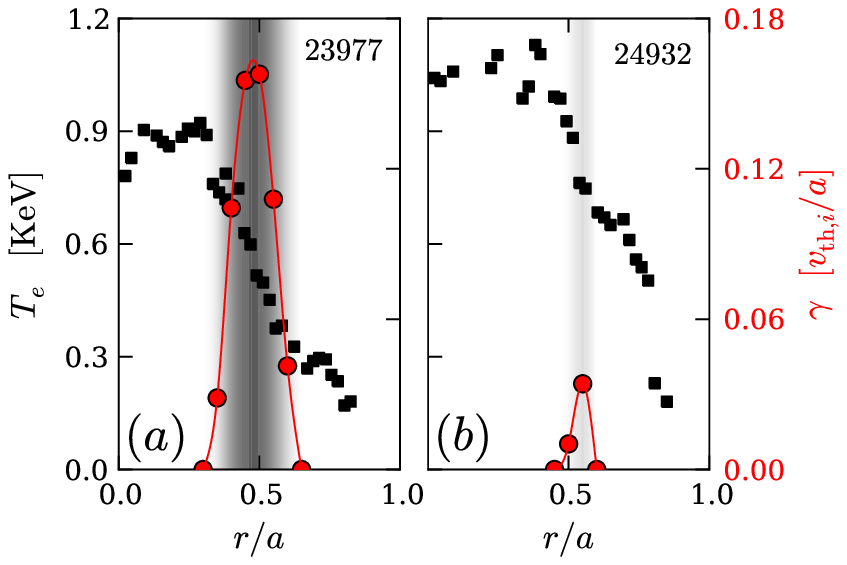}
\caption{(Color online) Electron temperature profiles (square symbols) for shot 23977 (a) and 24932 (b) at 64 ms, with the growth rate of the most unstable MT mode (circles); the gray-tone corresponds to the mode growth rate. The edge region $r/a\gtrsim 0.8$ is not considered in the present analysis.}
\label{fig:profiles}
\end{figure}

We present numerical investigations obtained with the gyrokinetic code GS2~\cite{gs2}, modified to include the RFP geometry~\cite{gs2itg}. Before discussing the results, we briefly introduce some technical aspects related to code and experiment. Simulations are linear, with fluctuations of the magnetic vector potential in both parallel and perpendicular components included. MT modes are characterized by a very elongated structure along the field lines; in order to generate spatially resolved eigenmodes, the simulation domain extends to $\theta\in[-80\,\pi,80\,\pi]$, where the poloidal angle $\theta$ plays the role of coordinate along the field line in the ballooning representation. The radial domain $r/a\le 0.8$ ($r\in[0,a]$, $a$ torus minor radius) is considered here, as we are interested in the ITB region. The geometry is assumed axisymmetric for simplicity, as described in~\cite{gs2itg}. The analysis is based on the profiles of two reference experimental cases selected from the database of SHAx discharges, i.e., shots 23977 and 24932, whose electron temperatures are shown in Fig.~\ref{fig:profiles}. The electron-to-ion temperature ratio is estimated to be $T_e/T_i\sim 1.5$, based on spectroscopic Doppler measurements of partially ionized impurities \cite{eps09}. Accordingly, the total $\beta$, ratio of thermal to magnetic pressure, which rules the electromagnetic effects in the gyrokinetic equation, is approximately 1.5\% in the region corresponding to the ITB for both reference discharges. The collision frequency and the electron diamagnetic frequency stay roughly in the range $\nu\sim 10\,\omega^*\sim 10^5$ s$^{-1}$.

\begin{figure}
\includegraphics{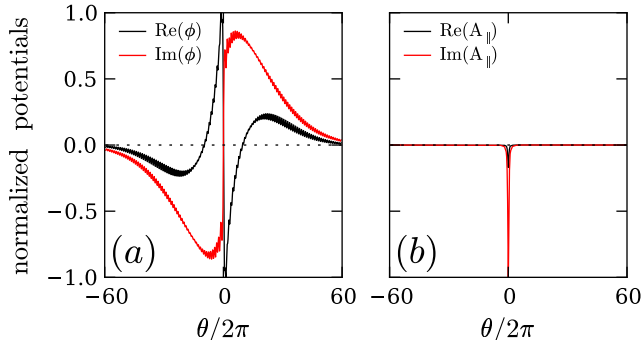}
\caption{(Color online) Normalized fluctuations of electrostatic potential (a) and parallel magnetic vector potential (b) for the profile of Fig.~\ref{fig:profiles}(a) at normalized radius $r/a=0.55$ and $\ky=0.18$.}
\label{fig:eigenmodes}
\end{figure}

\begin{figure}
\includegraphics{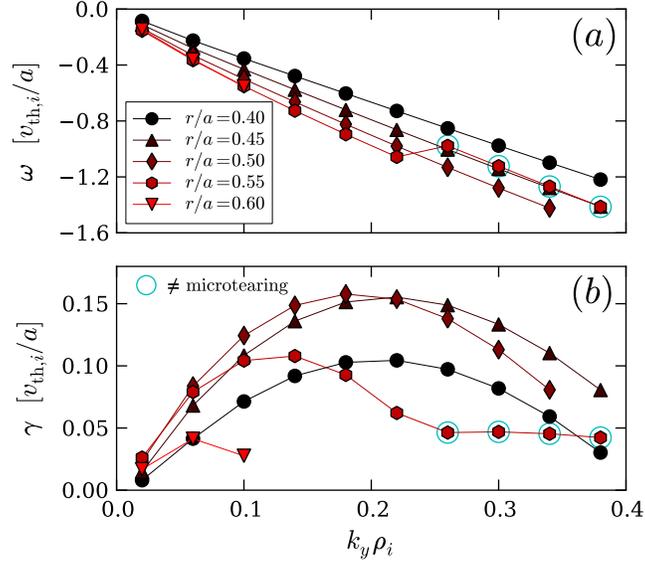}
\caption{(Color online) Mode frequency (a) and growth rate (b) of the most unstable mode as a function of the wavenumber $\ky$, for different mid-radius positions along the profile of Fig.~\ref{fig:profiles}(a). $\bigcirc$ symbols represent modes which do not have microtearing parity. Frequencies are normalized to $\vthi/a$, with $\vthi=(T_i/m_i)^{1/2}$ ion thermal speed and $a$ minor radius.}
\label{fig:spectra}
\end{figure}

Our study shows that the dominant core instability is of microtearing type, Fig.~\ref{fig:profiles}. Essentially driven by the higher electron temperature gradient, the discharge 23977 turns out to be more prone to this type of instability. The tearing nature of such instabilities is clear in Fig.~\ref{fig:eigenmodes}, where the typical parity of the mode, odd for the electrostatic potential $\phi$ and even for the parallel magnetic vector potential $A_\parallel$, is shown. In Fig.~\ref{fig:spectra} mode frequency and growth rate are plotted as a function of the perpendicular wavenumber $k_y$ multiplied by the thermal ion Larmor radius $\rho_i$, for different radial positions along the profile of Fig.~\ref{fig:profiles}(a). MT modes are characterized by a propagation in the electron diamagnetic direction (negative real frequency in the plot). The second branch in the spectrum for $r/a=0.55$ at $\ky>0.25$ corresponds to modes with even parity in $\phi$, characterized by a very elongated envelope of the eigenfunction in the parallel direction. Such modes resemble those described in \cite{hallatschek}, essentially determined by the passing-electron response.

\begin{figure}
\includegraphics{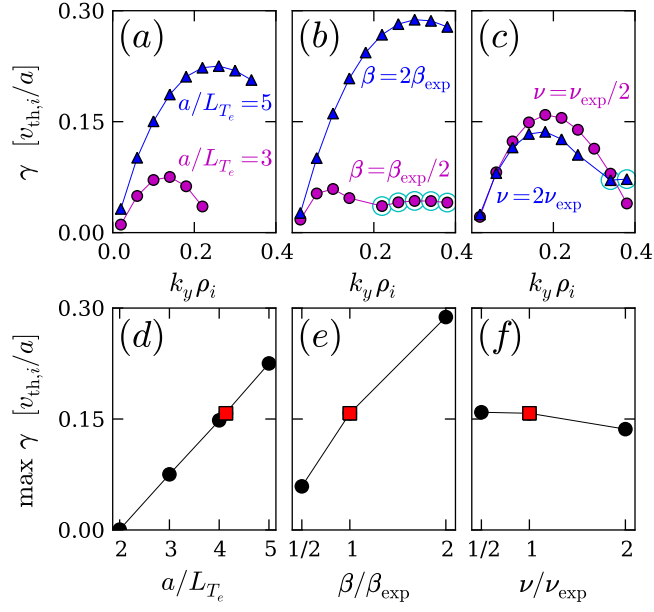}
\caption{(Color online) Dependence of the growth rate of the most unstable mode on $\ky$ for different values of $a/\lte$ (a), $\beta/\beta_{\rm exp}$ (b), and $\nu/\nu_{\rm exp}$ (c), the experimental parameters referring to the profile shown in Fig.~\ref{fig:profiles}(a) at $r/a=0.5$; $\bigcirc$ symbols represent modes which do not have microtearing parity. Frames (d), (e) and (f) summarize the dependence of the maximum of the above spectra versus the respective parameter, with the experimental conditions marked by square symbols.}
\label{fig:trends}
\end{figure}

To get a more complete picture of MT mode stability in the RFP, we performed a scan over a larger range of physical parameters. The MT mode growth rate is mainly affected by electron temperature gradient, collision frequency and total $\beta$. The reference values are assumed to be those of shot 23977 at $r/a=0.5$, see Fig.~\ref{fig:profiles}(a). All quantities are normalized to these experimental values, and the single parameters of interest are individually modified keeping the others fixed. The normalized logarithmic temperature gradient $a/\lte=-a\,d\log T_e/dr$ is varied in the interval $[2,5]$, the total $\beta$ and the collision frequency $\nu$ between one half and twice their respective experimental values. In Fig.~\ref{fig:trends}(a) the growth rate is shown as a function of $\ky$, for two values of $a/\lte$. The peak of the spectrum is moving towards larger $\ky$ as the gradient is increased, together with the maximum of the spectrum. Frame (d) summarizes the dependence of the maximum of the spectrum on $a/\lte$, showing a good linear trend and a stability threshold for $a/\lte\simeq 2$. Fig.~\ref{fig:trends}(b) deals with the $\beta$ parameter dependence. Large $\beta$ increases the growth rate and the wavenumber corresponding to the maximum growth rate. As $\beta$ decreases, MT modes tend to vanish, and a sub-dominant branch, peaked around $\ky\sim 0.3$ and not visible for the reference experimental $\beta$, is disclosed; again, such modes possess very extended tails along $\theta$ and qualitatively replicate those of \cite{hallatschek}. The maximum $\gamma$ versus $\beta$ shows a strong increasing trend, see frame (e). In view of the development of plasma density control and of the capability to achieve higher $\beta$, MT-driven turbulence could become a concern for ITB degradation. In Fig.~\ref{fig:trends}(c) the spectrum is reported for the two selected values of the collision frequency, with the maximum growth rate as a function of $\nu$ drawn in frame (f). As reported in~\cite{gladd80}, the growth rate of MT modes is expected to have a maximum for $\nu/\omega^*$ of the order of tens (depending on the value of $\eta_e=d\log T_e/d\log n$, with $n$ plasma density), and to decay logarithmically for larger frequencies, until stability is found. Similarly, a few additional cases more distant from the scan centered on the reference collisionality of Fig.~\ref{fig:trends}(c,f), indicate that $\gamma(\nu)$ features a maximum just around the experimental value of $\nu/\omega^*\sim10$.

We have established that MT modes are unstable over a significant range of wavenumbers, but substantial heat transport ensues only in the presence of appreciable field line stochastization. For this to happen, the Chirikov island overlap criterion adapted to the MT scenario~\cite{ippolito} yields that unstable modes must have poloidal numbers $m$ greater than the threshold $m_0 = q/(2|q'|\rho_i)^{1/2}$, with $q$ safety factor and $q'$ its radial derivative. This makes the minimum separation between mode rational surfaces, $\Delta r\simeq q^2/m^2|q'|$, less than a couple of ion Larmor radii, which is the typical width of the resistive layer. In our case $m_0\sim 3$ in the region where MT modes are unstable, whereas the peaks of the mid-radius spectra in Fig.~\ref{fig:spectra}(b) approximately correspond to $m\sim 10$, making the inequality $m>m_0$ fulfilled for a wide subset of wavenumbers.

\begin{figure}
\includegraphics{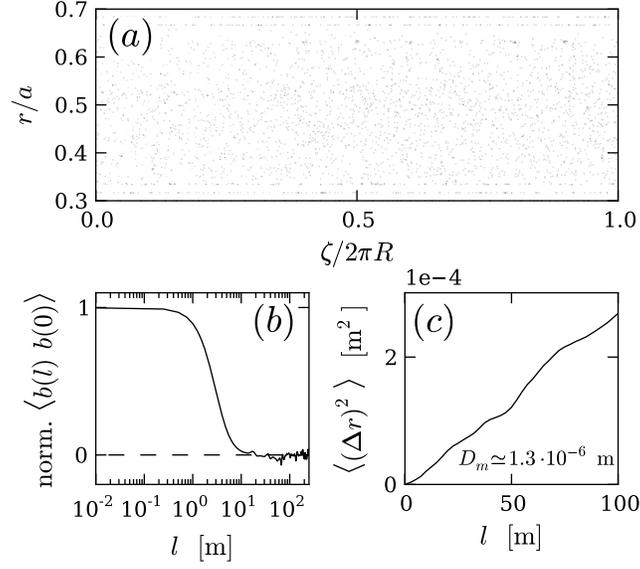}
\caption{Magnetic field line section at $\theta=\pi$, in frame (a), with corresponding longitudinal correlation function of the normalized radial field perturbation (b), and averaged squared radial displacement of the field lines as a function of the covered distance (c). Calculations of (b) and (c) are based on 2500 field line trajectories.}
\label{fig:nemato}
\end{figure}

To quantify the heat transport deriving from the saturated state of MT turbulence, the electron thermal conductivity $\chi$ must be estimated. Because of the difficulty involved in calculating such a quantity from direct nonlinear numerical simulations -- to our knowledge this has never actually been done for this kind of turbulence -- people commonly resort to the quasilinear estimate by Rechester and Rosenbluth~\cite{rr}. An upper limit for the conductivity is given by the collisionless estimate $\chi=\dmag\vthe$, where $\dmag$ is the magnetic diffusion coefficient, ratio of the averaged squared radial displacement to the distance traveled along the field lines, $\dmag=\langle(\Delta r)^2\rangle/2l$, and $\vthe$ is the electron thermal speed. The diffusion coefficient itself is usually written as $\dmag\sim b^2\lcorr$, with $b=\delta B/B$ normalized magnetic field perturbation and $\lcorr$ longitudinal correlation length of the field. The saturated level of the magnetic field perturbation is provided by the quasi-linear estimate by Drake \textit{et al}~\cite{gladd80}: $b\sim\rho_e/\lte$. In tokamaks $\lcorr$ is usually assumed approximately equal to the connection length~\cite{kadomtsev}. Due to the low value of $q$, for an RFP the connection length is expected to be much smaller, close to $2\pi a$ as the reversal is approached. To get a more accurate evaluation of $\lcorr$ we relied on numerical simulations with the field-line-tracing code \textsc{Nemato}~\cite{nemato}. The overall magnetic field is built by superposing a large population of modes with poloidal and toroidal wave-numbers $(m,n)$ determined by the spectra of Fig.~\ref{fig:spectra}, and with amplitude such that their cumulative strength is locally $\sim\rho_e/\lte$. The resulting magnetic field cross-section is shown in Fig.~\ref{fig:nemato}(a). The code allows us to compute the Eulerian spatial correlation $C(l)=\langle b(l)\,b(0)\rangle/\langle b^2(0)\rangle$, averaged over an ensemble of trajectories, as sketched in Fig.~\ref{fig:nemato}(b). The correlation length, given by $\lcorr=\int_0^\infty C(l)\,dl$, becomes constant for large values of the upper limit of integration, where $C(l)$ tightly oscillates around 0. It is confirmed that $\lcorr\approx 2\pi a$, this value turning out to be slightly decreasing for larger amplitudes of the magnetic field perturbation. In frame (c) we show the direct calculation of $\langle(\Delta r)^2\rangle$ as a function of $l$. The inferred numerical value of $\dmag$ agrees with the quantity $b^2\lcorr$. When the values of $\lcorr$, or directly the value of $\dmag$, are inserted into the expression for the quasilinear thermal conductivity, we obtain $\chi\sim 5\div 20\;{\rm m^2/s}$. This turns out to be compatible with the estimate for $\chi$ given at the barrier location by power balance analysis~\cite{rfx,fassina}, where one recovers figures in the range $5\div 50\;{\rm m^2/s}$. As a side remark, our evaluation of $\chi$ concerns the ITB MT-dominated region only, and does not take into account any possible mixing with the residual MHD turbulence acting in the edge of the plasma.

As a further support to MT turbulence as a major drive of transport across ITBs, we note that marginal stability of MT modes occurs at $a/\lte\gtrsim 2$, the precise figure depending upon a combination of several parameters [Fig.~\ref{fig:trends}(d)]. Consistently with the two sample cases addressed in this Letter, first statistical investigations on the RFX-mod database suggest that the experimental electron $\lte$ at the barrier typically exceeds the value $a/4$~\cite{rita}. A lower bound of $\lte$ is likely to be related to a gradient-driven turbulence mechanism which i) starts operating above a given threshold in the gradient $a/\lte$; ii) activates turbulent structures which increase local transport; iii) forbids reaching higher temperature gradients (hence the upper bound of $a/\lte$), due to the high cross-field thermal conductivity which tends to smooth the $T_e$ profile itself.

As a final remark, we notice that our findings are consistent with the results from spherical tokamaks~\cite{sphtok}, suggesting that a similar physics is running in the background. This is not at all surprising, as RFPs are acquiring a level of magnetic order until now deemed peculiar to tokamaks or stellarators.

To conclude, this work suggests that MT modes may contribute significantly to heat transport across ITBs in present-day RFX-mod plasmas. Our results are in good agreement with the experimental estimates of the thermal conductivity. It is still a matter of investigation how much MT modes may affect future RFP performances, with special attention paid to the scalings in $\beta$ and collisionality.

We are grateful to C. Angioni, S.~C. Guo, M.~E. Puiatti, R. Lorenzini, and D. Escande for very helpful discussions, and to M. Kotschenreuther and W. Dorland for making the code GS2 available. We are indebted to L.~Chacon for providing the code \textsc{Nemato}. We also thank S. Menmuir and M. Valisa for carefully revising the manuscript. This work has been supported by the European Communities under the contract of Association between Euratom/ENEA.



\end{document}